\documentstyle[12pt,psfig]{article}
\begin{document}
\begin{titlepage}
   \title{Schwarzschild Field  with
               Maximal Acceleration Corrections}
\author{A. Feoli$^{a, b}$, G. Lambiase$^a$,
G. Papini$^c$\thanks{E-mail: papini@cas.uregina.ca},
G. Scarpetta$^{a,d}$\thanks{E-mail: Scarpetta@physics.unisa.it}}
\date{\empty}
\maketitle
\centerline{\em  $^a$Dipartimento  di  Scienze Fisiche ``E.R. Caianiello'',  Universit\`a di Salerno, Italia.}

\centerline{\em $^a$Istituto Nazionale di Fisica Nucleare, Sez. di Napoli.}

\centerline{\em  $^b$Facolt\`a d'Ingegneria, Universit\`a del Sannio, Benevento, Italia}

\centerline{\em $^c$Dept. of Physics, University of Regina, Regina, Sask. S4S 0A2, Canada}

\centerline{\em $^d$ International Institute for Advanced Scientific Studies, Vietri sul Mare (SA), Italia}

\bigskip
\begin{abstract}

We consider a model in which accelerated particles experience line--elements
with maximal acceleration corrections. When applied to the Schwarzschild metric,
the effective field experienced by accelerated test particles contains
corrections that vanish in the limit $\hbar\to 0$, but otherwise affect the
behaviour of matter greatly. A new effect appears in the form of a spherical
shell, external to the Schwarzschild sphere, impenetrable to classical
particles.

\end{abstract}

\thispagestyle{empty}

\vspace{20.mm}


\vfill

\end{titlepage}

One of the main problems of modern theoretical physics is the
unification of Quantum Mechanics and General Relativity. A very interesting
step in this direction was taken by
Caianiello \cite{qg} who interpreted quantization  as curvature of the
relativistic eight dimensional space--time tangent bundle
$TM = M_{4}\otimes TM_{4}$, where $M_4$ is the usual flat space--time manifold
of metric $\eta_{\mu\nu}$ and signature -2.
In this space the standard operators of the Heisenberg algebra
are represented as covariant derivatives and
the quantum commutation relations are interpreted
 as components of the curvature tensor.
The Born reciprocity principle and, equivalently, the symmetry between
configuration and momentum space
representations of field theory, are
thus automatically satisfied.

An interesting feature
of Caianiello's model is that the proper accelerations
of massive particles along their worldlines
are normalized to an upper limit
${\cal A}_m$ \cite{ma},
referred to as maximal acceleration (MA).
 In some works  ${\cal A}_m$ is fixed by the Planck mass
$m_P = \left({\hbar c\over G}\right)^{1/2}$ \cite{b},\cite{infl} and is therefore a
universal constant. A direct application of Heisenberg's uncertainty relations
\cite{ca}, \cite{pw}, as well as the geometric interpretation of the
quantum commutation relations, suggest, however, that the natural limit
for the proper acceleration of any massive particle
 be fixed by the particle
rest mass itself according to the relation
${\cal A}_m=2m c^3/\hbar$.
 Its existence, which would
automatically rid  black hole entropy of ultraviolet divergences
\cite{BHEntropy},\cite{McG},
can  be also surmised in the
context of Weyl space \cite{pap} and of a geometrical
analogue of Vigier's stochastic theory \cite{jv}, or
conjectured on diverse classical and quantum grounds \cite{prove},\cite{b}.

The notion of MA is linked to the extended nature of  particles.
The inconsistency of the point--particle
concept for a relativistic quantum particle
is discussed by Hegerfeldt \cite{he}
who shows that
the localization of the particle at a given
point at a given time  enters in incurable
conflict with causality.

The introduction of an invariant interval in the eight-dimensional
space--time tangent fiber bundle $TM$, may  be also
interpreted as a regularization procedure of the field equations,
 alternative  to that in which space--time
is quantized by means of
a fundamental length, as  in \cite{qs},
where the two-point and four point
Green's functions of a massless scalar field theory are explicitely
constructed and shown to lead to the disappearance of
ultraviolet divergences at the one loop level.
{\it The advantage of Caianiello's proposal is to
preserve the continuum structure of space--time}.

MA has long been familiar in the classical context
\cite{wh}.

In a quantum relativistic context, the analysis of string
propagation in cosmological backgrounds has revealed that accelerations
higher than a critical value give rise to
Jeans--like instabilities \cite{gsv}.
These occur \cite{gasp} when the
acceleration induced by the background gravitational
field is large enough to render
the two string extremities causally disconnected because of the Rindler
horizon associated with their relative acceleration. The critical acceleration
$a_c$ is determined by the string  size $\lambda$ and is
given by $a_c = \lambda^{-1} =  (m\alpha)^{-1}$, where m is the string
mass and $\alpha^{-1}$ the usual string tension.

Frolov and Sanchez \cite{fs} have then found that a universal critical acceleration
$a_c \simeq \lambda^{-1}$ must be a general property of the strings.

In all these instances the critical acceleration is a consequence of
 the interplay of the Rindler horizon with the finite
extension of the particle.
In Caianiello's proposal the maximal proper acceleration
is a basic physical property of all massive particles, which
is an inescapable consequence of quantum mechanics \cite{ca},
\cite{pw},
and must therefore be included from the outset in the physical laws.
This requires a modification
of the metric structure of space-time.
It leads, in the case of Rindler space, to a manifold with a non-vanishing
scalar curvature and a shift in the horizon \cite{emb}. The cut-off on the
acceleration is the same required in an {\it ad hoc} fashion by
Sanchez in order to regularize the entropy and the free energy of quantum
strings \cite{sa2}. MA is also invoked as a necessary cut--off by
McGuigan in the calculation of black hole entropy \cite{McG}.

Applications of Caianiello's model include cosmology \cite{infl},
where the initial singularity can be avoided while preserving inflation,
the dynamics of accelerated strings \cite{Feo},
the energy spectrum of a uniformly accelerated particle \cite{emb},
the periodic structure as a function of momentum in neutrino
oscillations \cite{8}
and the expansion of the very early universe \cite{gasp}.
The model also makes the metric observer--dependent,
as conjectured by Gibbons and Hawking \cite{Haw}.

The extreme large value that ${\cal A}_m$ takes for all known particles
makes a direct test of the model very difficult. Nonetheless a direct test
that uses
photons in a cavity has also been suggested \cite{15}.
More recently, we have worked out the
consequences of the model for the classical electrodynamics of a particle \cite{cla},
the mass of the Higgs boson \cite{Higgs} and the Lamb shift in hydrogenic
atoms \cite{lamb}. In the last instance the agreement between experimental
data and MA corrections is very good for $H$ and $D$. For $He^+$ the
agreement between theory and experiment is improved by $50\%$ when
MA corrections are included. MA effects in muonic atoms
also appear to be measurable \cite{muo}

In this paper we analyze the modifications produced by MA
in the motion of a scalar particle in the  Schwarzschild field.

As stated above, there is indeed a simple way to
endow space-time with a causal structure
in which proper accelerations are limited. It consists in replacing
the usual Minkowski line element $ds^2=\eta_{\mu\nu}dx^{\mu}dx^\nu $
 with the
line element in the eight-dimensional
space-time tangent bundle $TM$
\begin{equation}\label{eq1}
d\tau^2=\eta_{AB}dX^AdX^B \qquad\qquad A, \, B = 0, \ldots, 7,
\end{equation}
where
$$
\eta_{AB}=\eta_{\mu\nu}\otimes \eta_{\mu\nu}\,{,}
$$
$$
X^{A}=\left(x^{\mu},\frac{c^2}{{\cal A}_m}
\frac{dx^{\mu}}{ds}\right)\qquad\qquad \mu=0,\ldots,3\,{,}
$$
$x^{\mu}=(ct,\vec{x})$ is the usual space-time four-vector and $dx^{\mu}/
{ds}=\dot x^{\mu}$ the four-velocity.

In order to write the
equations of motion in ordinary four--dimensional
space--time, an embedding procedure is required \cite{emb},
the first step of a process of successive
approximations,  which leads  to an effective four--dimensio\-nal space--time
geometry, induced on  $M_4$ by the metric $g_{A B}$ of $TM$ through
the parametric equations which govern the embedding of $M_4$ in $TM$.
Given the coordinates $\xi^\mu$, chosen to parametrize $M_4$, and
assigned on $M_4$ the velocity field $\dot x^\mu (\xi)$, the eight
equations $x^\mu = x^\mu (\xi)$ and $\dot x^\mu = \dot x^\mu (\xi)$
constitute the set of equations needed to represent $M_4$ as
an embedded submanifold of $TM$. The metric $g_{\mu \nu}(\xi)$,
locally induced, is then given by
$g_{\mu \nu}(\xi)= \eta_{\alpha\beta}\left({\partial x^\alpha\over
\partial \xi^\mu }{\partial x^\beta \over
\partial \xi^\nu}+ {1\over {\cal A}_m^2}{\partial \dot x^\alpha\over
\partial \xi^\mu }{\partial \dot x^\beta \over
\partial \xi^\nu}\right).$

The first order approximation introduced by this procedure
consists in defining
the velocity field $\dot x^\mu$
as the one obtained by solving the ordinary four--dimensional equations of motion.
The invariant line element (\ref{eq1}) can therefore be written
in the form
$$
d\tau^2=\eta_{\mu\nu}dx^{\mu}dx^\nu +{1\over {\cal A}_m^2}
\eta_{\mu\nu}d\dot x^{\mu}d\dot x^\nu=
$$
\begin{equation}\label{eq2}
= \left[1+\frac{\ddot x_\mu\ddot x^\mu}{{\cal A}_m^2}\right]
\eta_{\mu\nu} dx^\mu dx^\nu  {.}
\end{equation}
 As a consequence one obtains
mass-dependent corrections
to the effective space-time geometry experienced by accelerated
particles, which in general induce
curvature, and give rise to a mass dependent violation of
the equivalence principle.
In the classical limit
$\left ({\cal A}_m\right)^{-1} = {\hbar\over 2 m c^3}\rightarrow 0$
the terms contributing to the modification of the geometry vanish
and one returns to the ordinary space-time geometry.

In the presence of gravity, we replace  $\eta_{\mu\nu}$
with the corresponding metric tensor $g_{\mu\nu}$,
a  natural choice which preserves the full
structure introduced in the case of flat space.
We obtain
\begin{equation}\label{eq3}
d\tau^2=\left(1+\frac{g_{\mu\nu}\ddot{x}^{\mu}\ddot{x}^{\nu}}{{\cal A}_m^2}
\right)g_{\alpha\beta}dx^{\alpha}dx^{\beta}\equiv
\sigma^2(x) g_{\alpha\beta}dx^{\alpha}dx^{\beta}\,{.}
\end{equation}
The four--acceleration
$\ddot x^\mu = d^2 x^\mu/d\,s^2$ appearing in the conformal factor
of  (\ref{eq3}) is not a covariant quantity necessarily
orthogonal to the four--velocity $\dot x^\mu$, as in Minkowski space.
The justification for this choice lies primarily with the quantum mechanical
derivation of ${\cal A}_m$ \cite{ca},\cite{pw} which applies to $\ddot x^\mu$,
is Newtonian in spirit (it requires the notion of force) and is only
compatible with special
relativity. No extension of this derivation to general relativity
 has so far been given. The choice of $\ddot x^\mu$ in (\ref{eq3})
 is, of course, supported by the weak field approximation to $g_{\mu\nu}$
 which, to first order, is entirely Minkowskian.
Other relevant distinctions must be made. The model introduced is not
intended to supersede general relativity, but only to provide a
method to calculate the MA corrections to a Schwarzschild line
element. The effective gravitational field introduced in  (\ref{eq3})
can not be easily incorporated in general relativity (it violates,
for one, the
equivalence principle). Nor are the symmetries of general relativity
indiscriminately applicable to (\ref{eq3}). For instance, the conformal factor
is not an invariant, nor can it be eliminated by means of general
coordinate transformations. The embedding procedure requires that it
be present and that it be calculated in the same coordinates of the
unperturbed gravitational background. It is useful to keep these
distinctions in mind in what follows.

For convenience, the natural units $\hbar =c=G=1$ will be used below.

In order to calculate the corrections to the Schwarzschild field experienced
by a  particle initially at infinity and falling toward the origin
along a geodesic, one must calculate the metric induced by the embedding
procedure (\ref{eq2})
to first order in the parameter ${{\cal A}^{-2}_m}$.
On choosing $\theta =\pi/2$, one finds the conformal factor
produced by the embedding procedure
\begin{equation}\label{eq4}
\sigma^2(r)=1+\frac{1}{{\cal A}^2_m}
\left[\left(1-\frac{2M}{r}\right)\ddot{t}^2-
\frac{\ddot{r}^2}{1-2M/r}-r^2\ddot{\phi}^2\right]\,{,}
\end{equation}
where $\ddot{t},\ddot{r}$ and $\ddot{\phi}$ are given by the standard
results \cite{wh}
\begin{eqnarray}
\ddot{t}^2 & = &\frac{\tilde{E}^2}{(1-2M/r)^4}\frac{4M^2}{r^4}
\left[\tilde{E}^2-\left(1-\frac{2M}{r}\right)\left(1+\frac{\tilde{L}^2}{r^2}
\right)\right]\,{,} \nonumber \\
\ddot{r}^2 & = &\left(-\frac{M}{r^2}+\frac{\tilde{L}^2}{r^3}-
\frac{3M\tilde{L}^2}{r^4}\right)^2\,{,}\label{eq5} \\
\ddot{\phi}^2 & = & \frac{4\tilde{L}^2}{r^6}\left[\tilde{E}^2-
\left(1-\frac{2M}{r}\right)\left(1+\frac{\tilde{L}^2}{r^2}
\right)\right]\,{.} \nonumber
\end{eqnarray}
$M$ is the mass of the source, $\tilde{E}$ and $\tilde{L}$ are the total
energy and angular momentum per unit of  particle mass $m$.
The conformal factor $\sigma^2(r)$ is then given by
$$
\sigma^2(r)=1+\frac{1}{{\cal A}_m^2}\left\{
-\frac{1}{1-2M/r}\left(-\frac{3M\tilde{L}^2}{r^4}+\frac{\tilde{L}^2}{r^3}
-\frac{M}{r^2}\right)^2 + \right.
$$
\begin{equation}\label{eq6}
\left.
+\left(-\frac{4\tilde{L}^2}{r^4}+\frac{4\tilde{E}^2 M^2}{r^4(1-2M/r)^3}
\right)\left[\tilde{E}^2-\left(1-\frac{2M}{r}\right)\left(1+
\frac{\tilde{L}^2}{r^2}\right)\right]\right\}\,{.}
\end{equation}
In order to make a direct comparison with the motion in Schwarzschild geometry
possible, we adopt the same procedure and notations of Ch. 25 of \cite{wh}.
The only exception is here represented by our choice of metric signature
$(-2)$. Starting from the fundamental statement that the magnitude of the
energy--momentum four--vector is given by the rest mass $m$ of the particle $
g_{\alpha\beta}p^{\alpha}p^{\beta}=m^2$
and using the modified metric (\ref{eq3}), one
obtains
\begin{equation}\label{eq7}
\left(\frac{dr}{d\tau}\right)^2=\frac{1}{\sigma^4(r)}
\left[\tilde{E}^2-\left(1-\frac{2M}{r}\right)\left(\sigma^2(r)+
\frac{\tilde{L}^2}{r^2}\right)\right]\equiv \tilde{E}^2-
\tilde{V}^2_{eff}(r)\,{,}
\end{equation}
which corresponds to the equation
\begin{equation}\label{eq8}
\left(\frac{dr}{d\tau}\right)^2=
\tilde{E}^2-\left(1-\frac{2M}{r}\right)\left(1+
\frac{\tilde{L}^2}{r^2}\right)\equiv \tilde{E}^2-\tilde{V}^2(r)\,{,}
\end{equation}
of Ref. \cite{wh}. The analysis of the motion can therefore be
given in term of
\begin{equation}\label{eq9}
\tilde{V}^2_{eff}(r)=\tilde{E}^2-\frac{\tilde{E}^2}{\sigma^4(r)}+
\frac{1}{\sigma^2(r)}\left(1-\frac{2M}{r}\right)\left(1+\frac{\tilde{L}^2}
{r^2\sigma^2(r)}\right)\,{.}
\end{equation}
In a more traditional approach one would write
\begin{equation}\label{eq10}
\frac{1}{2}\left(\frac{dr}{d\tau}\right)^2+\tilde{P}_{eff}(r)=\omega\,{,}
\end{equation}
where $\omega\equiv (\tilde{E}^2-1)/2$ \cite{wh}.
Substituting (\ref{eq8}) into (\ref{eq10}) one obtains $
\tilde{P}_{eff}=\frac{1}{2}\, \left(\tilde{V}^2_{eff}(r)-1\right)${.}
It now is convenient
to introduce the adimensional variable $\rho=r/M$ and the parameters
$\epsilon=(M{\cal A}_m)^{-1}$
and $\lambda =\tilde{L}/M$.
The complete expression for $\tilde{V}^2_{eff}(\rho)$ is then
\begin{equation}\label{eq11}
\tilde{V}^2_{eff}(\rho)=\tilde{E}^2\left\{1+\frac{1}{\sigma^2(\rho)}
\left[-\frac{1}{\sigma^2(\rho)}+\frac{1}{\tilde{E}^2}\left(1-\frac{2}{\rho}\right)\left(
1+\frac{\lambda^2}{\rho^2\sigma^2(\rho)}\right)\right]\right\}\,{.}
\end{equation}

Notice that $\tilde{V}^2_{eff}\to \tilde{E}^2$ as $\rho \to 2$ and
$\rho \to 0$. Moreover
$\tilde{V}^2_{eff}\to 1$ as $\rho \to \infty$.
Plots of (\ref{eq11})  for different values of $\tilde{E}$ show
a characteristic step--like behaviour in the neighborhood of $\rho =2$
(Fig. 1 for  $\lambda = 0$, radial motion,  Fig. 2 for $\lambda \ne 0$).
Fig. 1 clearly shows that the effective
and Schwarzschild potentials cannot be distinguished
from each other for distances ranging from infinity to points very near
the Schwarzschild radius, where
$\tilde{V}^2_{eff}$
acquires a marked dependence on the energy
of the  particle and develops a barrier which prevents the particle from reaching
the  Schwarzschild horizon. An expansion of (\ref{eq9})
 in the neighborhood of $\rho =2$ yields, in fact,
the behaviour of the height of the potential barrier as
$\tilde{V}^2_{eff}\sim\tilde{E}^2+\frac{(\rho -2)^4}{4\epsilon^2\tilde{E}^4}
+O((\rho -2)^5)$
which, clearly, has the minimum   $\tilde{E}^2$  on the horizon $\rho =2$.
This term vanishes only in the limit $\tilde{E}\to \infty$ and/or
in the limit $\epsilon\to \infty$ for
which ${\cal A}_m$ or $M$ or both vanish and the problem becomes meaningless.

 A second interesting consequence of the line element (\ref{eq3}) is
 represented by the shift to the left in
the horizon position in the case of radial motion, $\lambda = 0$ (Fig. 1). The new horizon actually
becomes a true singularity. Nevertheless,
the potential barrier prevents incoming massive particles of energy $\tilde{E}$ from falling
into the pit in the effective potential. The addition of MA
effects does therefore produce, already to lowest order in ${\cal A}_m^{-2}$,
a spherical shell of radius $2<\rho <2+\eta$, with  $\eta \ll 1$
where, according to (\ref{eq7}) and (\ref{eq8}),
the velocity of any incoming  particle becomes imaginary.  Such a shell is
classically impenetrable and remains so at higher orders of approximation
in ${\cal A}_m^{-2}$.  The analogous occurrence of a classically impenetrable shell
was derived by Gasperini as a
consequence of the breaking of the local $SO(3,1)$ symmetry \cite{MG}. A generalization of the Schwarzschild
metric to include MA corrections was also given in \cite{8} for the two-dimensional problem of particles in
hyperbolic motion in a Kruskal plane. As is known, these particles are static relative to Schwarzschild
coordinates ($r = const$). The classically impenetrable shell and shift in horizon also occur in this instance.

Fig. 2 shows that the scenario is similar to the previous case for particles
with energy $\tilde{E}$ higher than the centrifugal
potential barrier (dot--dashed line). If, on the contrary,
 the particle's energy  is
lower, its motion does not differ from the classical one
because in the physically accessible region  (${\tilde V}^2 \le {\tilde E}^2$)
$\tilde{V}^2_{eff}$ does not differ
  significantly  from the Schwarzschild potential. The modifications induced
in the region near the Schwarzschild radius, i.e. the infinite potential barrier for
low values of ${\tilde E}^2$ (dashed line in Fig. 2), or the finite, but very
high potential barrier for intermediate values of ${\tilde E}^2$ (dotted line in Fig. 2),
have no influence on the  particle motion, because they occur in a region
precluded by the centrifugal barrier. The existence of bound orbits is assured by the fact that
$\tilde{V}_{eff}$ and the Schwarzschild potentials reach the same
value rapidly for $\rho > 2$.

Fig 3 provides information about the localization of the singularities of the effective
potential, according to the values of ${\tilde E}$ and ${\lambda}$. For radial motion
(dashed line in fig. 3) and for increasing values of ${\tilde E}$ one meets one, than three
and than again one singularity point, always located in the  region internal to the  event horizon.
 When the normalized angular momentum ${\lambda}$ differs from zero
 (solid line in fig. 3), there
is a critical value of ${\tilde E}$ above which no singularities appear. Below this
critical value,
 two singularity points appear on the left of the Schwarzschild radius. At
 even lower energies
the singularities become four, two on the left and two on the right of
the Schwarzschild radius.

The singularities can best be analyzed by calculating the scalar curvature $R(\rho) \equiv R_\mu^\mu$
and the Kretschmann curvature invariant
 $ I \equiv R_{\alpha \beta \gamma \delta} R^{\alpha \beta \gamma \delta}$ for the metric
(\ref{eq3}). The explicit results are cumbersome and will be given elsewhere.
We have however ascertained that
the $\lim_{\epsilon\to 0}R(\rho)=0$, as it should,  because for $\epsilon =0$
(\ref{eq3}) yields the unmodified Schwarzschild metric.
 Moreover
$\lim_{\rho\to 2}\, R(\rho) = 0$,
$\lim_{\rho\to 0}\, R(\rho)=0$ for $\lambda\ne 0$,
$\lim_{\rho\to 0}\, R(\rho)=-54{\cal A}_m^{2}$ for $\lambda = 0$ indicating
that  there is no singularity at the origin.
The behaviour of $M^2 R(\rho)$ for
$\epsilon=0.001$ and different values of   $\lambda$ and $\tilde{E}$ is given in Fig. 4 for
$\rho\ge 2$.

The metric modifications (\ref{eq3})
discussed in the previous section give rise to a
 new, interesting effect represented by the presence of a spherical
shell, external to the Schwarzschild sphere, that is forbidden to any
classical  particle. Its implications are clear: no massive particle can
reach the Schwarzschild horizon.
Would, however, such a shell prevent the
formation of a black hole? Certainly yes if the model is correct and
accretion of massive particles is a viable process of formation for black holes.
One must however be cautious and keep in mind that the occurrence of the shell is essentially
dynamical and that the procedure followed is perturbative.

Additional questions must also be considered.

i) {\it Massive quantum tunneling through the shell $2<\rho < 2+\eta$}.
Preliminary calculations based on spinless particles and the Klein--Gordon
equation do not seem to favour this possibility, but a more detailed
analysis is necessary before reaching definitive conclusions.

ii) {\it Inflow at constant speed}. One may assume that accumulation of
matter in proximity of the shell would provide a viscous background to the
motion of other incoming or outgoing matter. Particles moving with constant
speed would experience a normal Schwarzschild field and fall into the
pit  for suitable values of
$\tilde{L}$ and $\tilde{E}$. It may be difficult to realize this situation
in proximity of a strong source in any realistic way, but, a priori, not
impossible.

iii) {\it Trasformation of matter into photons}. The existence of large
accelerations ($\ddot{x}_{\mu}\ddot{x}^{\mu}\sim 0.02{\cal A}_m^2$) in proximity of the
shell would generate enormous streams of photons (and gravitons)
by brehmsstrahlung. Substantial capture of photons by a large gravitational
field would in time lead to black hole formation because massless particles
see unmodified Schwarzschild fields.
In this case formation would be only delayed and the whole process would
resemble a gigantic  engine of which photons
(and possibly other massless particles) would at the same time be
product and fuel.

The power radiated away by a single electron in proximity of the shell
can be estimated from the formula \cite{LAN}, \cite{cla}
$P=-(2e^2/3c)\ddot{x}_{\mu}\ddot{x}^{\mu}$.
An expansion of $\ddot{x}_{\mu}\ddot{x}^{\mu}$ in the neighborhood $r=2M+\delta M$ of
the shell yields $\ddot{x}_{\mu}\ddot{x}^{\mu}\sim -\delta^4 {\cal A}_m^{-2}/(4{\tilde E}^4)+O(\delta^5)$,
and typically $\delta\sim
0.005$. The power emitted is therefore independet of the mass of the source.

For $\tilde{E}=3$ one finds $P\sim 23.9erg/s$. At the highest accelerations,
$\ddot{x}_{\mu}\ddot{x}^{\mu}\sim 10^{-2}{\cal A}_m^2$, $P\sim 3.1\cdot 10^{8}erg/s$,
and the conversion rest mass energy into radiation
becomes extremely efficient.

\bigskip
\bigskip

\begin{centerline}
{\bf Acknowledgments}
\end{centerline}

Research supported by NATO Collaborative
Research Grant No. 970150, by Italian Ministero dell'Universit\`a e della
Ricerca Scientifica, fund ex 40\% and 60\% DPR 382/80,
and by the Natural Sciences and Engineering Research Council of Canada.


\bigskip

\bigskip

\begin{thebibliography}{99}

\bibitem{qg} E.R. Caianiello, {\it La Rivista del Nuovo Cimento}
              {\bf 15} n.4 (1992) and references therein.
\bibitem{ma}  E.R. Caianiello, {\it Lett. Nuovo Cimento} {\bf 32} (1981) 65;
              E.R. Caianiello, S. De Filippo, G. Marmo, and G.Vilasi,
                    {\it Lett. Nuovo Cimento} {\bf 34} (1982) 112.
\bibitem{b}     H.E. Brandt, {\it Lett. Nuovo   Cimento} {\bf 38}
                (1983) 522; {\bf 39} (1984) 192;{\it Found. Phys. Lett.} {\bf 2} (1989) 39;
                {\bf 4} (1991) 523; {\bf 11} (1998) 265.
\bibitem{infl}  E.R. Caianiello, M. Gasperini, G. Scarpetta,
                {\it Class. Quantum Grav.} {\bf 8} (1991) 659; \\
                M. Gasperini, {\it in} ``Advances in Theoretical Physics''
                ed. E.R. Caianiello, (World Scientific, Singapore, 1991),
            p. 77.
\bibitem{ca}    E.R. Caianiello, {\it Lett. Nuovo Cimento} {\bf 41} (1984) 370.
\bibitem{pw}    W.R. Wood, G. Papini and Y.Q. Cai, {\it Il Nuovo Cimento}
                {\bf B104}, 361 and (errata corrige) 727 (1989).
\bibitem{BHEntropy} G. `t Hooft, {\it Nucl. Phys.} {\bf B256} (1985) 727;\\
                L.Susskind, J. Uglam, {\it Phys. Rev. } {\bf D50} (1994)
                2700.
\bibitem{McG}   M. McGuigan, {\it Phys. Rev. } {\bf D50} (1994) 5225.
\bibitem{pap} G. Papini and W.R. Wood, {\it Phys. Lett.} {\bf A170} (1992) 409;
              W.R. Wood and G. Papini, {\it Phys. Rev.} {\bf D45} (1992) 3617;
              {\it Found. Phys. Lett.} {\bf 6} (1993) 409;
              G. Papini, {\it Mathematica Japonica} {\bf 41} (1995) 81.
\bibitem{jv}  J. P. Vigier, Found. Phys. 21 (1991) 125.
\bibitem{prove} A. Das, {\it J. Math. Phys.} {\bf 21} (1980) 1506; \\
                M. Gasperini, {\it  Astrophys. Space Sci} {\bf 138}
                         (1987) 387; \\
                M. Toller, {\it Nuovo Cimento} {\bf B102} (1988) 261;
            {\it Int. J. Theor. Phys.} {\bf 29} (1990) 963;
                {\it Phys. Lett.} {\bf B256} (1991) 215; \\
                R. Parentani, R. Potting, {\it Phys. Rev. Lett.} {\bf 63} (1989)
            945;\\
            P. Voracek, {\it  Astrophys. Space Sci} {\bf 159}
                         (1989) 181; \\
                B.   Mashhoon, {\it  Physics  Letters}{\bf A143} (1990) 176; \\
                V. de Sabbata, C. Sivaram, {\it Astrophys. Space Sci.}
                {\bf 176} (1991) 145;
                {\em Spin and Torsion in gravitation}, World Scientific,
                Singapore, (1994);\\
                D.F. Falla, P.T. Landsberg, {\it Nuovo Cimento } {\bf B 106},
                (1991) 669;\\
                A.K. Pati, {\it Nuovo Cimento} {\bf B 107} (1992) 895;
                {\it Europhys. Lett. }{\bf 18} (1992) 285.
\bibitem{he}    G.C. Hegerfeldt, Phys. Rev. {\bf 10 D} (1974) 3320.
\bibitem{qs}    J.C. Breckenridge, V. Elias, T.G. Steele,
                {\it Class. Quantum Grav.}{\bf 12} (1995) 637.

\bibitem{wh}    C.W. Misner, K.S.Thorne, J. A. Wheeler, {\it Gravitation},
                W.H. Freeman and Company, S. Francisco, 1973.
\bibitem{gsv}   N. Sanchez and G. Veneziano, {\it Nucl. Phys}. {\bf 333 B},
                       (1990) 253; \\
                M. Gasperini, N. Sanchez, G. Veneziano, {\it Nucl. Phys.},
            {\bf 364 B} (1991) 365; {\it Int. J. Mod. Phys}.{\bf 6 A}
                       (1991) 3853.
\bibitem{gasp}  M. Gasperini, {\it Phys. Lett}. {\bf258 B} (1991) 70;
            {\it Gen. Rel. Grav.} {\bf 24} (1992) 219.
\bibitem{fs}    V.P. Frolov and  N. Sanchez. {\it Nucl. Phys.}
                   {\bf 349 B} (1991) 815.
\bibitem{emb}   E.R. Caianiello, A. Feoli, M. Gasperini, G. Scarpetta,
                {\it Int. J. Theor. Phys.} {\bf 29} (1990) 131.
\bibitem{sa2}    N. Sanchez, in {\it ``Structure: from Physics to General
                Systems''} eds. M. Marinaro and G. Scarpetta
                (World Scientific, Singapore, 1993) vol. 1, pag. 118.
\bibitem{Feo}   A. Feoli, {\it Nucl. Phys.} {\bf B396} (1993) 261.
\bibitem{8}     E.R. Caianiello, M. Gasperini, and G. Scarpetta,
                {\it Il Nuovo Cimento} {\bf B105} (1990) 259.
\bibitem{Haw}  G. Gibbons and S.W. Hawking, {Phys. Rev.} {\bf D15} (1977) 2738.
\bibitem{15}  G. Papini, A. Feoli, and G. Scarpetta, {\it Phys, Lett.}
              {\bf A202} (1995) 50.
\bibitem{cla} A. Feoli, G. Lambiase, G. Papini, and G. Scarpetta, {\it Nuovo Cimento}
              {\bf 112B} (1997) 913.
\bibitem{Higgs} G. Lambiase, G. Papini, and G. Scarpetta, {\it Nuovo Cimento}
              {\bf 114B} (1999) 189. See also
                S. Kuwata, {\it Il Nuovo Cimento} {\bf 111B} (1996) 893.
\bibitem{lamb}  G. Lambiase, G. Papini, and G. Scarpetta, {\it Phys. Lett.}
                {\bf 244A} (1998) 349.
\bibitem{muo}   C.X. Chen, G. Papini, N. Mobed, G. Lambiase and G. Scarpetta,
                {\it Nuovo Cimento}
              {\bf 114B} (1999) 199.
\bibitem{LAN}  L.D. Landau and E.M. Lifshitz, {\it The Classical Theory of
               Fields} (Pergamon Press, Oxford) 1969.
\bibitem{MG} M. Gasperini, {\it Phys. Rev.} {\bf D34} (1986) 2260.

\end{thebibliography}
\end{document}